\documentclass[%
 reprint,
 amsmath,amssymb,
 aps,
]{revtex4-1}
\usepackage{graphicx}% Include figure files
\usepackage{dcolumn}% Align table columns on decimal point
\usepackage{bm}% bold math
\usepackage{color}
\begin{document}

\preprint{APS/123-QED}

\title{Generation of coherent two-color pulses at the two adjacent harmonics in a seeded free-electron laser}% Force line breaks with \\
\author{Zhouyu Zhao}
 %\altaffiliation{liheting@ustc.edu.cn.}%Lines break automatically or can be forced with \\corresponding author:
\author{Heting Li}%
 \email{liheting@ustc.edu.cn}
 \author{Qika Jia}
\affiliation{%
National Synchrotron Radiation Laboratory, University of Science and Technology of China, Hefei, 230029, Anhui, P.R.China\\
}%



\begin{abstract}

The growing requirements of pump-probe techniques and nonlinear optics experiments greatly promote the studies of two-color free-electron lasers (FELs). We propose a new method to generate coherent two-color pulses in a high-gain harmonic generation (HGHG) FEL. In this scheme, an initial tilted electron beam is sent though the modulator and dispersive section of an HGHG FEL to generate the bunching at harmonics of the seed laser. Then a transverse gradient undulator (TGU) is adopted as the radiator and in such radiator, only two separated fractions of the tilted beam will resonate at two adjacent harmonics of the seed laser and are enabled to emit the coherent two-color pulses simultaneously. The time separation between the two pulses are on the order of hundreds of femtoseconds, and can be precisely controlled by varying the tilted amplitude of the electron beam and/or the transverse gradient of the TGU radiator. Numerical simulations confirm the validity and feasibility of this scheme in the EUV waveband.

\begin{description}
\item[PACS numbers]
41.60.Cr, 41.85.Ct
\end{description}
\end{abstract}

\maketitle

%\tableofcontents

\section{Introduction}

The short-wavelength free-electron laser (FEL) characterized by high brightness, good coherence and ultra-short pulse promotes cutting-edge science in atomic physics, chemistry and biology, etc. For some specific experiments, the output characteristics of FEL pulses both in spectral and temporal domain are required to be adjusted to satisfy the experimental needs. One of the most important formats is to create double FEL pulses which contain two different spectral lines with adjustable time separation, namely, the so-called two-color pulses. Especially, the pump-probe experiments can be accessed by using two-color FEL pulses which open the door for scientists to study the structural dynamics on the atomic and molecular scale.

Recently, the multi-color FEL experiments with different methods have been proposed or carried out in various kinds of radiation frequency band. In summary, there are mainly several schemes to lase at different wavelength, including: (1) Two electron beams with different energy pass through an undulator and resonate at two different wavelengths, as studied at SPARC \cite{lab1, lab2, lab3, lab4, lab5}. In this approach, each pulse is independent then the saturation power are similar in condition of maintaining minimum undulator length. However, it is difficult to obtain such beam and the tunability of pulses both in spectral and temporal domain is limited. (2) A single mono-energetic electron beam passes through alternating undulator segments with different undulator fields \cite{lab6, lab7, lab8, lab9, lab10, lab11}. It could be able to adjust the pulse separation by changing magnetic chicane. However, normally, due to the two-color pulses are generated by the same beam, its saturated power is comparatively lower than in first approach. (3) A twins-pulse-seed laser is adopted to generate two-color FEL radiation with precisely controlled time separation and wavelength, which has been demonstrated at the FERMI FEL, one of the most promise two-color FEL sources \cite{lab12, lab13, lab14}. (4) A new method named the fresh-slice technique has been proposed recently \cite{lab15, lab16, lab17}. The beam goes through alternating undulator segments and then laser is generated by different part of the beam. By manipulating the vertical positions of the tail and head of the beam, this approach has the virtue of the central wavelength of each pulse could be adjusted independently, and the pulse separation could be also adjusted precisely. It allows further flexibility for generating two-color FEL pulses.

Inspired by these works, in this paper we propose to utilize a transverse tilted beam to create two-color FEL radiation based on the high-gain harmonic generation (HGHG) FEL with a TGU radiator. In this method, as shown in Fig. 1, a tilted beam is firstly modulated by a single seed laser and then dispersed by a magnetic chicane. As a result, bunching at harmonics is generated. In the TGU radiator, the electrons at different transverse position will experience different magnetic field and only when the electron energy and the magnetic field resonate at the harmonic of the seed laser, the intense radiation will be generated. Therefore, with the optimization of the tilted amplitude of the electron beam and the gradient of the TGU radiator, this scheme can be contributed to generate two-color pulses at the adjacent harmonics, e.g., one pulse at the 4th harmonic and another at the 5th harmonic. The detailed description of the working principle is given in Section 2 and three-dimension simulations are shown in the Section 3. Several issues are discussed in the next section and finally we summarize in the last section.

\begin{figure}[!hbt]
\includegraphics[height=4.5cm,width=8cm]{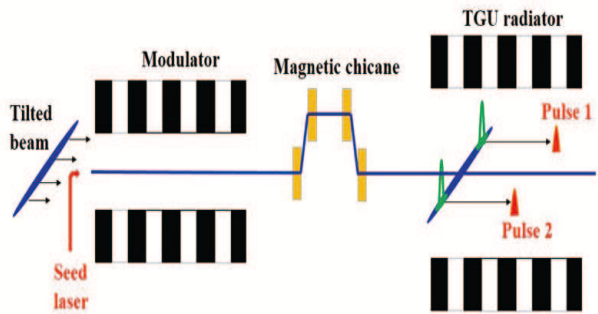}
\caption{Schematic of the proposed scheme for the two-color FEL generation.}
\label{fig:1}
\end{figure}

\section{Methods}

The electrons in the so-called 'tilted beam' can be characterized by the transverse offset $x$ in proportion to the longitudinal position $s$. The coupling between the longitudinal and transverse positions of the particles is normally unwanted in the conventional FEL facilities for the reason that only a fraction of the beam can contribute to the FEL radiation and therefore the corresponding output performance is degraded. However, the recent researches demonstrated that the tilted beam has many important applications, such as production of broad-bandwidth radiation \cite{lab18, lab19}, generation of ultrashort FEL pulse \cite{lab20, lab21, lab22} and generation of two-color pulses \cite{lab23, lab24} and so on. Such applications can satisfy some specific experimental requirements, for instance, the pump-probe experiments and nonlinear optics experiments. To obtain such a tilted beam with a small energy chirp, several methods have been developed, including the RF deflecting cavity \cite{lab25, lab26}, transverse wakefields in the accelerating structure and corrugated structure \cite{lab27, lab28, lab29}, residual dispersion from compression structure \cite{lab30, lab31}.

We assume an ideal tilted beam produced by the methods mentioned above with a linear tilted amplitude $\eta = dx/ds$ and an energy chirp on the order of 2\%. With such energy spread, the tilted amplitude can be tuned in a quite large range. As shown in Fig. 1, the tilted beam travels parallel along the normal modulator axis in which a seed laser is used to modulate the beam. The seed laser radius should be large enough to ensure the adequate energy modulation in each part of the beam.

After the beam passing through a magnetic chicane, the energy modulation is converted into density modulation and bunching at the harmonics is generated.
Since an energy chirp is introduced in the electron bunch, the dispersion on the whole bunch should be considered. The dispersive strength of the chicane in an HGHG FEL usually is not large, typically $R_{56}$ on the order of a few hundred micrometers. Considering a total energy chirp of 2\%, the variation of the whole bunch length is on the order of a few micrometers, which can be ignored compared with the initial bunch length.

In a conventional HGHG FEL, then the electron beam and the field of the radiator will resonate at the $n$th harmonic of the seed laser. The radiation wavelength $\lambda_1$ can be written as
\begin{eqnarray}
\lambda_1 = \frac{{\lambda_u}}{{2\gamma}^{2}}(1+\frac{K^2}{2})=\frac{\lambda_0}{n}
\end{eqnarray}
where $\gamma$ is the Lorentz factor of the electrons, $\lambda_0$ is the wavelength of the seed laser, and $K$, $\lambda_u$ are the strength parameter and period of the radiator, respectively.

However, in the proposed scheme, a transverse gradient undulator (TGU) is adopted as the radiator whose normalized field parameter $K$ has a linear $x$ dependence \cite{lab32, lab33},
\begin{eqnarray}
K(x) = K_0 (1+ \alpha x)
\end{eqnarray}
where $K_0$ is the dimensionless field strength on the axis of the TGU and $\alpha$ is the field gradient.

 In this TGU radiator, different part of the electron beam will experience different undulator magnetic field. The resonant wavelength $\lambda_r (s)$ along the whole bunch is

\begin{eqnarray}
\lambda_r (s) = \frac{{\lambda_u}}{{2\gamma}^{2}}(1+\frac{{K^2(x)}}{{2}})
\end{eqnarray}
One can see that the resonant wavelength from one side to another side (i.e., from the head to the tail) of the beam will become longer or shorter in a quadratic curve.

However, only the radiation at the harmonics of the seed laser will be coherently amplified because the electron beam is bunched at these discrete wavelengths.
If we carefully optimize the values of $\eta$, $K_0$ and $\alpha$ to make the resonant wavelengths along the whole bunch covering two adjacent harmonics of the seed laser, two short FEL pulses with different frequencies will be generated from two fractions of the bunch. Due to the limited gain bandwidth, each pulse will also have a short pulse length. The other electrons will experience self-amplified spontaneous emission and contribute little to the total output power.

The time separation between the two FEL pulses is determined by the longitudinal positions of the two bunch fractions that emit the coherent radiation. Combining the coupling relation $x = \eta s$ and Eq. (2), we can get
\begin{eqnarray}
\Delta t =\frac{\Delta s}{c}= \frac{\Delta K}{c \alpha \eta K_0}
\end{eqnarray}
where $c$ is the speed of the light in the vacuum.

Now we assume $\lambda_1$ and $\lambda_2$ are the radiation wavelengths at the two adjacent harmonics, corresponding to the $n$th and $(n+1)$th harmonic, respectively. From the resonance condition, we obtain

\begin{eqnarray}
\Delta K \approx \frac{{2\gamma}^{2}}{{\lambda_u}}\frac{1}{K_{aver}}\frac{\lambda_0}{n(n+1)}
\end{eqnarray}
Here $K_{aver}$ is the average $K$ value for the two radiation wavelengths. It is worth pointing out that we neglect the energy chirp for simplicity since the relative variation of the electron energy with the transverse position is much smaller than that of the undulator strength.

According to Eq. (5), the harmonic order $n$ should not be too small, otherwise the required gradient of the TGU will be too large. In contrast, due to the limited harmonic up-conversion efficiency, the harmonic order also can not be too large. Moreover, large harmonic orders will require a long
 TGU and lead to a low radiation power. Therefore, the harmonic order $n$ should be moderate.

To make the two-color pulses close to the axis, one should set the $K_{aver}$ as the undulator strength on the axis of the TGU radiator, namely, $K_0=K_{aver}$. On this condition, the two FEL pulses will locate on both sides of the undulator axis with an equal distance to the axis. Inserting Eq. (5) into Eq. (4), we achieve the time separation between the two FEL pulses as

\begin{eqnarray}
\Delta t \approx  \frac{1}{c \alpha \eta K_0^2}\frac{{2\gamma}^{2}}{{\lambda_u}} \frac{\lambda_0}{n(n+1)}
\end{eqnarray}
Obviously, the time separation is inversely proportional to $\alpha$ and $\eta$. And based on the relation of $x = \eta s$, the transverse separation of the two pulses can be easily given as
\begin{eqnarray}
\Delta x \approx  \frac{1}{\alpha K_0^2}\frac{{2\gamma}^{2}}{{\lambda_u}} \frac{\lambda_0}{n(n+1)}
\end{eqnarray}

\section{Simulations}
 To validate the feasibility of this new method, we have done the three-dimension time-dependent simulations based on the main parameters of Dalian Coherent Light Source (DCLS), which is an extreme ultraviolet (EUV) FEL user facility based on the principle of HGHG-FEL \cite{lab34}. Genesis 1.3 code \cite{lab35} was used to perform the two-color FEL generation.

  The electron beam used in the simulations has the following properties: a flattop beam with beam energy of 400 MeV, slice energy spread of 20 keV, normalized transverse emittance of 1.0 mm$\cdot$mrad, peak current of 1 kA and full width of 1 ps. In addition, a total energy chirp of 2\% and a linear tilted amplitude $\eta =11.7$ are introduced.

The modulator consists of 20 periods with period length of 50 mm.
A 240 nm seed laser with beam waist of 1.5 mm is used to modulate the electron beam.
Due to the gaussian distribution of the seed power in the transverse direction, the modulation is non-uniform for the electrons in different transverse positions.
The average modulation amplitude is $A=10$, where $A$ means the ratio of maximum energy modulation to the initial local energy spread.

Next, a magnetic chicane follows the modulator to convert the energy modulation into density modulation and then bunching at the harmonics of the seed laser is generated.
Meanwhile, the magnetic field will compress or decompress the whole electron bunch as there is a linear energy chirp.
In this simulation example, the dispersive strength is $R_{56}=0.21 $ mm and as a result,
the whole bunch length will vary by about 5 $\mu$m which can be neglected.

Then, the bunched beam enters the TGU radiator to emit coherent radiation at the harmonics of the seed laser.
We consider to generate the two-color pulses at the 4th and 5th harmonics of the seed laser, i.e., $n=4$ for Eq. (5).
The TGU radiator is composed of 50 periods with period length of 30 mm and transverse gradient of $\alpha=150$ m$^{-1}$.
Such short length of the radiator used here benefits from the high-quality electron beam and low harmonic orders converted from the seed laser.

\begin{figure}
\includegraphics[width=6.5cm]{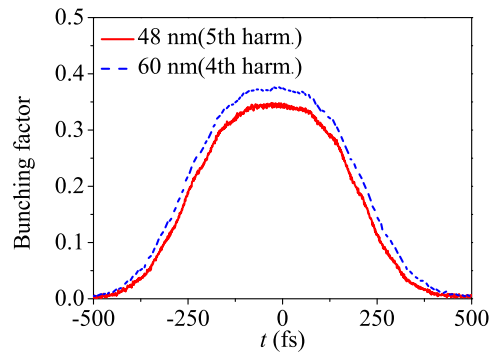}
\caption{The 4th and 5th harmonics' bunching factors of the electron beam at the exit of the dispersive section.}
\label{fig:2}
\end{figure}

The bunching factors for these two harmonics at the entrance of the radiator are shown in Fig. 2.
Obviously, the bunching factors also have gaussian shapes due to the non-uniform energy modulation.
To make the two fractions generating the two FEL pulses with considerable bunching factors, as mentioned in the previous section, we set the undulator parameter on the axis to be $K_0=K_{aver}= 1.65$, while the resonant undulator strengths for the 4th and 5th harmonics are $K=$1.83 and 1.48, respectively.

The time structure and spectrum of the output FEL pulses at the end of the TGU radiator are given in Fig. 3. For the 48 nm pulse (5th harmonic), the peak power is about 170 MW, the FWHM pulse width is 51 fs and the FWHM bandwidth is about 0.23\%, while for the 60 nm pulse (4th harmonic), the peak power is about 260 MW, the FWHM pulse width is 63 fs and the FWHM bandwidth is about 0.21\%.
Both the two pulses are close to Fourier transform limit and the radiation power of the two pulses is very close to each other. Owning to the lower bunching factor and higher harmonic orders, the power of 48 nm radiation is a little lower than that of 60 nm.
The time separation between the two pulse is about 386 fs while it is 345 fs calculated from Eq. (6).
\begin{figure}
\includegraphics[width=8.5cm]{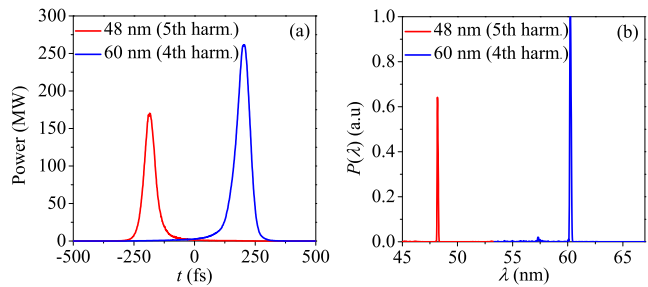}
\caption{The time structure (a) and the spectrum (b) of the two-color FEL pulses at the exit of the TGU radiator.}
\label{fig:3}
\end{figure}

\begin{figure}
\includegraphics[width=7cm]{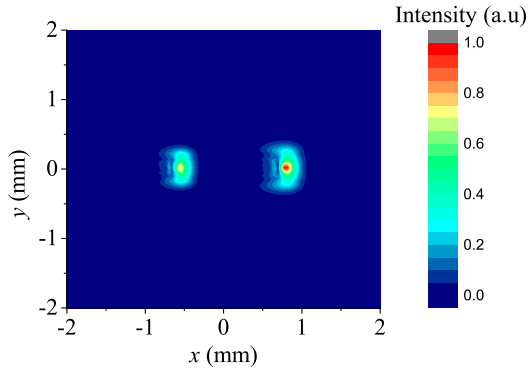}
\caption{The transverse distribution of the two-color pulses.}
\label{fig:4}
\end{figure}

Figure 4 shows the transverse distribution of the two-color pulses. It can be seen that the two pulses are almost symmetric about the undulator axis and have a transverse separation of about 1.4 mm  while it is 1.2 mm calculated from Eq. (7).

Note that, since the separation of the radiation wavelength is too large to be covered in a single run, the simulations of the FEL radiation in the radiator were done with two separate runs and each run focused on one harmonic of the seed laser. As the two FEL pulses are generated from two separated fractions of the electron bunch,
these two radiation processes have little impact on each other, therefore we think the simulation results are believable.

\section{Discussions}
In the proposed scheme, the generation of the two-color pulses relies upon the tilted beam and the transverse gradient field of the radiator.
Concerning these two important factors, several issues should be considered.

\begin{figure}
\includegraphics[width=6.5cm]{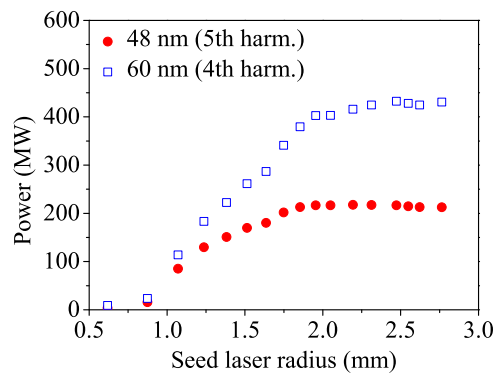}
\caption{The FEL power as a function of the rms seed laser radius. The average energy modulation amplitude is fixed to be $A=10$.}
\label{fig:5}
\end{figure}

 \begin{figure}
\includegraphics[width=8.5cm]{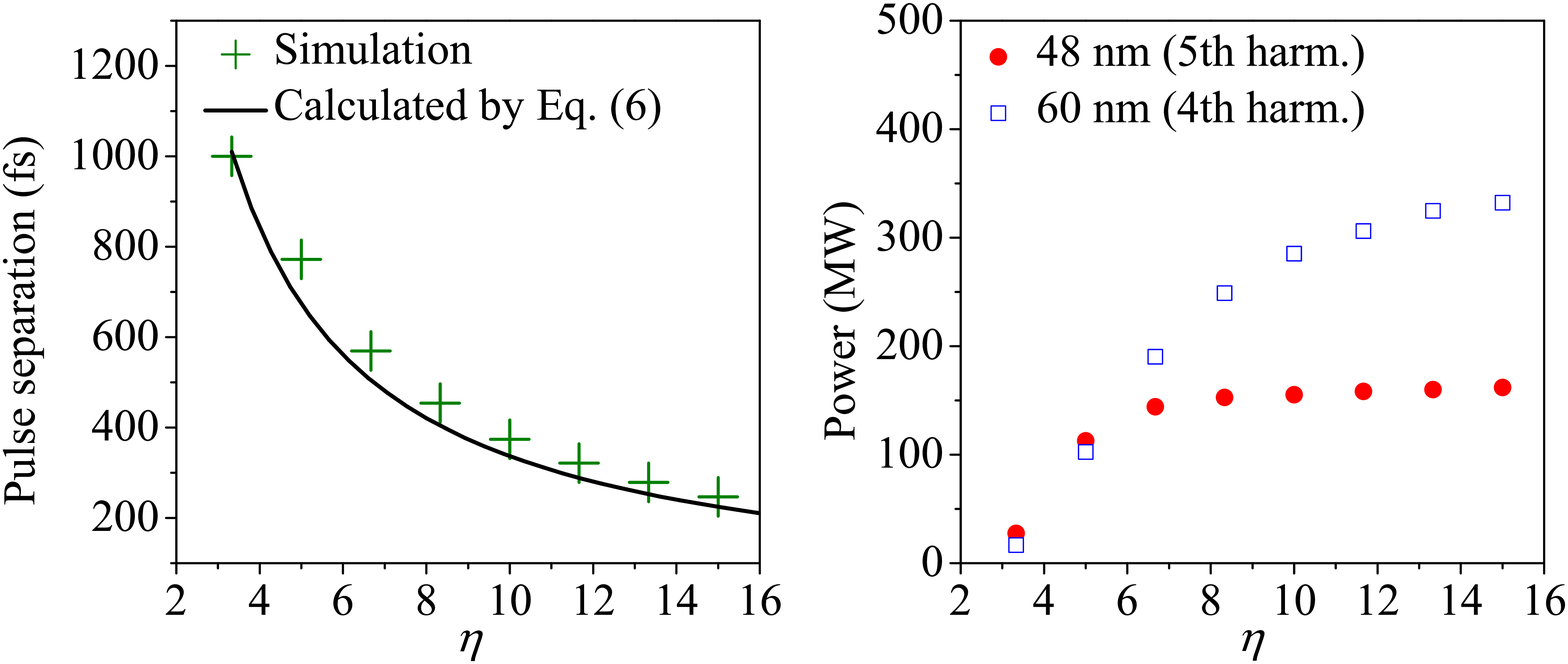}
\caption{The variations of the time separation (a) and the radiation power (b) of the two-color pulses with the tilted amplitude $\eta$. The transverse gradient is $\alpha=150$ m$^{-1}$.}
\end{figure}

\begin{figure}
\includegraphics[width=8.5cm]{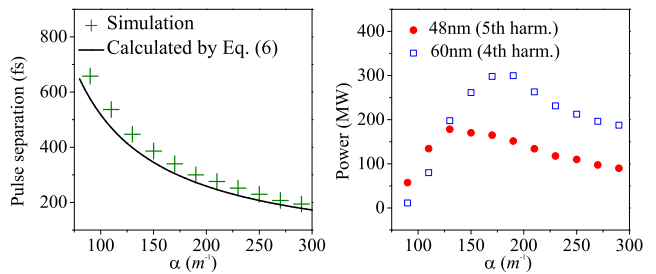}
\caption{The variations of the time separation (a) and the radiation power (b) of the two-color pulses with the transverse gradient $\alpha$. The tilted amplitude is $\eta = 11.7$.}
\label{fig:7}
\end{figure}

The initial tilted beam has a large horizontal size comparing with that in a normal FEL, therefore, a large transverse beam size of the seed laser is required. Considering that the two bunch fractions emitting the two FEL pulses locate in the two sides of the undulator axis symmetrically, the rms transverse beam size $\sigma _s$ of the seed laser should be at least larger than the half of the transverse distance of the two pulses, i.e., $\sigma _s>\Delta x/2$. Figure 5 shows the FEL power of the two pulses as a function of the rms radius of the seed laser. The average energy modulation amplitude is fixed to be $A=10$. When $\sigma _s<\Delta x/2$, the power of the two pulses is very low, and when $\sigma _s$ grows to 3 times larger than $\Delta x/2$, the powers of the two pulses do not grow any more with the further increase of the seed radius. However, as given in Eq. (6), the time separation of the two FEL pulses is not affected by the beam size of the seed laser.

Then we consider the tolerances of time separation and the radiation power on the tilted amplitude $\eta$ and the transverse gradient $\alpha$.
Figure 6 shows the variations of the time separation and the radiation power of the two-color pulses with the tilted amplitude of the electron beam and the variations with the transverse gradient are given in Fig. 7.

 From Fig. 6(a) and Fig. 7(a), one can find that the time separation is in inverse proportion to the tilted amplitude and the transverse gradient, and the simulation results are in good agreement with Eq. (6). Changing the tilted amplitude is more efficient for changing the time separation, because actually the transverse gradient is not easy to be changed for an existing TGU \cite{lab36}, but the tilted amplitude of the electron beam can be conveniently tuned by the methods mentioned in Section 2.
In addition, for a fixed bunch length, increasing the tilted amplitude $\eta$  or the transverse gradient $\alpha$ can make the width of each FEL pulse shorter.
From Fig. 6(b) and Fig. 7(b), it is clear that the radiation power is not sensitive to both of the tilted amplitude and the transverse gradient.

%\begin{figure}
%\includegraphics[width=8.5cm]{8.eps}
%\caption{The variations of the time separation with (a) the radiation power (b) of the two-color pulses with the transverse gradient $\alpha$. The tilted amplitude is $\eta = 11.7$.}
%\label{fig:7}
%\end{figure}

\section{Summary}
In summary we have proposed a new method to implement the two-color FEL operation in an HGHG FEL. A initial tilted electron beam and a TGU radiator are required in this scheme.
The simulation results based on DCLS parameters have shown that two coherent FEL pulses at the adjacent harmonics of the seed laser can be generated simultaneously from two separated fractions of the electron bunch in the TGU radiator.
These two pulses have a time separation of several hundred femtoseconds and a transverse distance of a few millimeters, and each pulse has a ultrashort pulse width of several ten femtoseconds with a bandwidth close to Fourier limit.
Both of the temporal and transverse separation can be precisely controlled by adjusting the tilted amplitude of the initial electron beam and/or the transverse gradient of the radiator.
Such magnitude of the time separation can directly meet the demand of some of the two-color user experiments.

 This scheme has the potential of generating a single ultrashort pulse, i.e., making only one fraction of the electron bunch resonate at one harmonic of the seed laser.
 For this case, a larger tilted amplitude and a lower harmonic order will make the single FEL pulse shorter. It also can be used to generate multicolor pulses, however, a large
transverse gradient will be required and may be very challenging.

\section{ACKNOWLEDGMENTS}
This work is supported by the National Natural Science Foundation of China under Grant (No. 11375199, 11205156), and National Key Research and Development Program of China (No. 2016YFA0401901).

\nocite{*}

\bibliography{}% Produces the bibliography via BibTeX.

\end{document}